\documentclass[a4paper]{article}  
\usepackage[a4paper, total={6in, 8in}]{geometry}
\usepackage{graphicx,cite}
\usepackage{subfigure}
\usepackage{amsmath}
\usepackage{float}

\begin{document}

\title{Temperature dependent quantum correlations in three dipolar coupled two-level atoms}
\author{Shaik Ahmed\\
School of Physics, University of Hyderabad\\ Hyderabad - 500046, India}
\maketitle  
\begin{abstract}

 \textbf{Abstract}: We investigate the thermal entanglement characteristics of three dipole-coupled two-level atoms arranged in two different configurations - in a line with nearest neighbour coupling and in a closed loop with each atom interacting with both its neighbours.  It is observed that in loop configuration, any one of the three atoms is indeed entangled with the other two atoms in the system, which are not mutually entangled, and further that this feature is specific to only the loop configuration, which is markedly absent in the line configuration.  A detailed study of  the quantum correlations demonstrated how these can be tuned by varying the temperature and the dipole dipole coupling strength, in both the configurations. 
\end{abstract}

\section{Introduction}
Entanglement, well-known for several decades\cite{Ein, Sch}, is one of the most striking features of quantum mechanics and plays a key role in quantum computation and quantum information\cite{Dik,Ekert,CH,Nielson} and is very sensitive to the system-environment interaction and the initial state of the system.  If a composite system is not entangled, normally one concludes that it is separable.  However, a composite system may contain other types of non-classical correlation, even when  it is separable.  The most popular measure of such correlations is the quantum discord introduced by Zurek et al.\cite{Olliver,Zurek,Henderson,Vedral}.  In comparison with entanglement and other
thermodynamic quantities, discord has received significant interest from the perspective of quantum phase transitions\cite{wer} and NMR quantum computations\cite{Auc}.  Recently, experimental estimation of quantum discord in a spin $1/2$ anti-ferromagnetic Heisenberg system was studied from the macroscopic properties such as susceptibility and heat capacity\cite{chiran}.  Many theoretical studies in interacting quantum spin systems have been performed to calculate pairwise correlations\cite{Dill,Werlang}.
 It is well known that a knowledge of different measures of entanglement and its distribution over each of the constituent atoms is crucial for tasks such as quantum teleportation\cite{Bennett} and super dense coding\cite{Bose, Ban, Pati}. 
  
In this paper, we investigate pairwise thermal entanglement characteristics of three atoms interacting through dipole-dipole coupling.  Further, we allow for the possibility of the three atoms to be aligned in a line as well as in a loop configuration.  Here, the entanglement property of the system is characterized by concurrence\cite{r4,r5} and the quantum correlations are characterized by quantum discord\cite{Olliver,Zurek,Henderson,Vedral}.  
In this work, we will show that the entanglement characteristics of three atoms undergo an unexpected, qualitative change when the atoms are arranged in loop configuration, which feature is markedly absent in the line configuration.  In particular, we observe an interesting result that any one of the qubits is indeed entangled with the other two qubits in the system, which are not mutually entangled, for any arbitrary dipole-dipole coupling strength.  We also show that the quantum correlations may be enhanced upon decreasing the temperature and increased with an increase in the dipole dipole coupling strength, both in line and loop configurations.
 
 The organization of the paper is as follows.
 In sec.II, we introduce our model and give the Hamiltonian for the system of three identical two-level atoms interacting with each other via dipole-dipole coupling. In Sec.III, we present numerical results for the pairwise entanglement characteristics of three two-level atoms arranged in line configuration.  In particular, we study the effects of temperature, atomic transition frequency and dipole-dipole coupling strength on pairwise concurrence and quantum discord.  We present the corresponding results for the atoms arranged in a loop - configuration, in Sec.IV.  Conclusions drawn from this study are presented in Section V, wherein we provide direction for further investigation.
 
\section{Formulation}
The Hamiltonian for the system of three atoms (A, B, C) coupled through dipole-dipole interaction is given by
 \begin{equation*}\label{eq:Hamiltonian}
H={\omega \sum_{i=A, B, C}}S^{z}_{i}+\sum_{i\neq j=A, B, C}\Omega_{ij}(S^{+}_{i}S^{-}_{j}+H.C) 
\end{equation*}
The first term describes the unperturbed energy of the system while the second term represents the dipole - dipole interaction  between the atoms, where  $ \Omega_{ij} $, the dipole dipole interaction strength, is a function of the inter-atomic separation `$ d $'. 
In the above, $ \omega $ is the atomic transition frequency,  $ S^{+}_{i} = |1\rangle_{i} \langle 0| $  and $S^{-}_{i}=|0 \rangle_{i} \langle 1| $  are the raising and lowering operators of the $i^{th}$ atom. 
Here, we assume that all the three atoms are identical, i.e., $ \omega_{A}=\omega_{B}=\omega_{C}=\omega $ and dipole dipole interaction strength $ \Omega_{ij} = \Omega $ for $ i\neq j= A, B, C$.
\begin{figure}[H]
\centerline{\includegraphics[width=12.0 cm,height=7.0 cm]{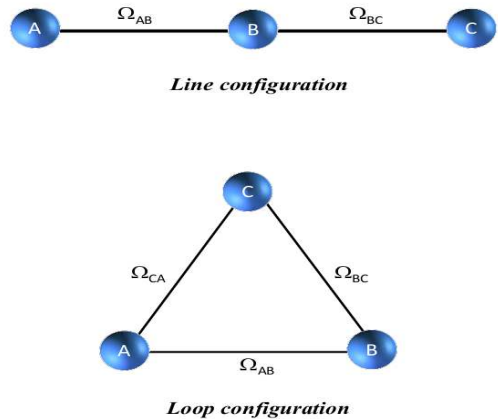}}
\caption[]{(Color online) Schematic representation of line and loop for three atoms coupled via dipole-dipole interaction}
\label{fig:One}
\end{figure}
 In the absence of any radiation field, the Hamiltonian (eq.\ref{eq:Hamiltonian}) is expanded in pure standard basis, $ | ijk \rangle $, where $ i,j,k=0,1 $, by diagonalizing which all the eigenstates will be obtained.  
  
In this work, we consider two of the configurations, line and loop, in which the system of three atoms can be arranged.  In the line configuration, the atoms are located in a linear array, wherein the d-d interactions
 between successive atoms in this linear chain are considered, i.e., the interaction between atoms A and B ($\Omega_{AB}$), and that between atoms B and C ($\Omega_{BC}$) alone is taken into account.  The second configuration we consider is a closed loop, where each of the three atoms interacts with its neighbours.  Both these configurations are shown in figure\ref{fig:One}.  We are considering these two configurations which are distinct and it should be noted that  any other arrangement would be a simple variation of these two.  As the atom-atom coupling manifests in different ways in both these configurations, the results for the two configurations are  found to be markedly different. 

 For the three atom system, in thermal equilibrium at temperature T, the density matrix can be expressed as 
\begin{equation}\label{eq:rhoABC}
\rho_{ABC}={\frac{1}{Z}}\sum^{8}_{i=1} |\psi_{i} \rangle \langle \psi_{i}| e^{-\beta \epsilon_{i}} ,
\end{equation}
where $Z=\mathrm{Tr}(e^{-\beta \epsilon_{i}})$ is the partition function and $ \beta=\frac{1}{k_{B}T} $ ($ k_{B}$
is Boltzmann constant).  Given how crucial the measure of entanglement and its distribution over different atoms is, for tasks such as quantum teleportation\cite{Bennett} and super dense coding\cite{Bose, Ban, Pati}, in this work we  investigate the entanglement properties of the three atoms system.  

 The focus here is on the pairwise entanglement associated with any two atoms in terms of concurrence $ C(\rho_{ij}) $ and quantum discord $ D(\rho_{ij}) $.    
 For pairwise thermal entanglement, one can obtain the reduced density matrices $ \rho_{ij} $ by taking partial transpose of $ \rho_{ijk} $ with respect to $ k $, given by 
\begin{equation}
\rho_{ij}= \mathrm{Tr}_{k}(\rho_{ijk}) 
\end{equation} 
where $ i,j,k=A,B,C $.  In the following, we give a brief description of the quantum discord.
    
\subsection{Quantum Discord}
 For two correlated subsystems A and B, quantum discord is defined as the difference between the total correlation and the classical correlation, given by the following expression $ D=I-Q $.
 Here the total correlation between two subsystems A and B of a bipartite quantum system ($ \rho_{AB} $) can be quantified by the quantum mutual information $I$\cite{rev,Berry}, 
\begin{equation}
I(\rho_{AB})=S(\rho_{A})+S(\rho_{B})-S(\rho_{AB})
\end{equation}
  where $ \rho_{A} $ and $ \rho_{B} $ are the reduced density matrices for subsystems A and B and $S (\rho)(\rho = {\rho_{A}, \rho_{B}, \rho_{AB}})$ represents the von Neumann entropy,
\begin{equation}
S(\rho)=-\mathrm{Tr}\rho \log \rho .
\end{equation}
 The measurements performed on one system, in general, influence the state of the other system, even if the two systems are far away from each other and do not directly interact\cite{Holevo}.  Postulating that the total classical part of correlations is the maximal amount of information about one subsystem, say A, that can be extracted by performing a measurement on the other subsystem B, Henderson and Vedral\cite{Henderson} suggested taking, as a measure of classical correlation, the quantity
\begin{equation}
C(\rho_{AB})=max_{\lbrace B_{i}\rbrace }\lbrace S(\rho_{A})-\sum_{i}p_{i}S(\rho^i_{A})\rbrace.
\end{equation}
Here $ \lbrace B_{i}\rbrace $ is a complete set of measurements on the subsystem B and $ \rho^i_{A} $
is the remaining state of A after obtaining the outcome $ i $ on B, and $ p_{i} $
is the probability to detect the result $ i $.
Now, the quantum discord $ D(\rho_{AB}) $ is given by the difference between quantum mutual information $ I(\rho_{AB}) $ and the measurement-induced quantum mutual information $ C(\rho_{AB})$
\begin{equation}
D(\rho_{AB})=I(\rho_{AB})-C(\rho_{AB})
\end{equation}
 and it is interpreted as a measure of quantum correlations\cite{Olliver}.  In what follows, we present details of the quantum correlations in both line and loop configurations. 

\section[Line configuration]{The Quantum correlations in Line configuration}

At thermal equilibrium, the quantum state of a three atom system is a weighted superposition of all the eigenstates.   By diagonalizing the Hamiltonian H (Eq.\ref{eq:Hamiltonian}), we can obtain all the eigenvalues $ \epsilon_{i} $and their corresponding eigenstates $|\psi_{i} \rangle $.  The eigenvalues $\epsilon_i$, in the line configuration, are
\begin{align}
 \epsilon_{1}&=\frac{-3\omega}{2};~\epsilon_{2}=-\sqrt{2}\Omega-\frac{\omega}{2};~\epsilon_{3}=-\frac{\omega}{2};~ \epsilon_{4}=\sqrt{2}\Omega-\frac{\omega}{2}\nonumber \\
\epsilon_{5}&=-\sqrt{2}\Omega+\frac{\omega}{2};~\epsilon_{6}=\frac{\omega}{2}; ~ \epsilon_{7}=\sqrt{2}\Omega+\frac{\omega}{2};~\epsilon_{8}=\frac{3\omega}{2} 
\end{align}
and the corresponding  eigenstates $|\psi_{i} \rangle $ of the system, are given by
\begin{align}
 |\psi_{1} \rangle &= |g_{1}g_{2}g_{3} \rangle ; |\psi_{2} \rangle = \frac{1}{2}\left[|e_{1}g_{2}g_{3} \rangle -\sqrt{2}|g_{1}e_{2}g_{3} \rangle +|g_{1}g_{2}e_{3}\rangle \right]\nonumber \\
 |\psi_{3} \rangle &= \frac{1}{\sqrt{2}}\Big[|g_{1}g_{2}e_{3} \rangle - |e_{1}g_{2}g_{3} \rangle \Big]; |\psi_{4} \rangle = \frac{1}{2}\left[|e_{1}g_{2}g_{3} \rangle +\sqrt{2}|g_{1}e_{2}g_{3}  \rangle +|g_{1}g_{2}e_{3} \rangle \right]\nonumber \\
 |\psi_{5} \rangle &= \frac{1}{2}\left[|e_{1}e_{2}g_{3} \rangle -\sqrt{2}|e_{1}g_{2}e_{3} \rangle +|g_{1}e_{2}e_{3}  \rangle \right]; |\psi_{6} \rangle = \frac{1}{\sqrt{2}}\Big[|g_{1}e_{2}e_{3} \rangle - |e_{1}e_{2}g_{3} \rangle \Big]  \nonumber \\
 |\psi_{7} \rangle &= \frac{1}{2}\left[|e_{1}e_{2}g_{3} \rangle +\sqrt{2}|e_{1}g_{2}e_{3} \rangle +|g_{1}e_{2}e_{3}  \rangle \right]; |\psi_{8}\rangle = |e_{1}e_{2}e_{3} \rangle. \nonumber \\
 \end{align}

Therefore, considering the standard basis \{$|g_{1}g_{2}g_{3} \rangle $, $|e_{1}g_{2}g_{3} \rangle $, $|g_{1}e_{2}g_{3} \rangle $, $ |g_{1}g_{2}e_{3}\rangle $, $|e_{1}e_{2}g_{3} \rangle $, $|e_{1}g_{2}e_{3} \rangle $, $|e_{1}e_{2}g_{3} \rangle $ , $|e_{1}e_{2}e_{3}\rangle$\} of the system, one can obtain the thermal density matrix of the form
\begin{equation}
  \rho_{ABC}(T)=\dfrac{1}{Z}\begin{bmatrix}
\rho_{11} & 0 & 0 & 0 & 0  & 0 & 0 & 0 & \\
0 & \rho_{22} & \rho_{23} & \rho_{24} & 0  & 0  & 0 & 0  \\
0 & \rho_{32} & \rho_{33} & \rho_{34} & 0 & 0 & 0 & 0  \\
0 & \rho_{42} & \rho_{43} & \rho_{44} & 0 & 0 & 0 &  0  \\
0 & 0 & 0& 0 & \rho_{55} & \rho_{56} & \rho_{57} & 0 \\
0 & 0 & 0 & 0 & \rho_{65} & \rho_{66} & \rho_{67}  & 0 \\
0 & 0 & 0 & 0 & \rho_{75} & \rho_{76} & \rho_{77} & 0\\
0 & 0 & 0  & 0 & 0 & 0 & 0 & \rho_{88} \\
\end{bmatrix} 
\label{line_density}
\end{equation}  
where the partition function
$$ Z=e^{\frac{3\omega}{2k_{B}T}}+e^{\frac{-3\omega}{2k_{B}T}}+ e^{\frac{(\frac{\omega}{2}-\sqrt{2}\Omega)}{k_{B}T}}+e^{-\frac{(\frac{\omega}{2}+\sqrt{2}\Omega)}{k_{B}T}}+e^{\frac{(\frac{\omega}{2}+\sqrt{2}\Omega)}{k_{B}T}}+e^{-\frac{(\frac{\omega}{2}-\sqrt{2}\Omega)}{k_{B}T}}+e^{\frac{\omega}{2k_{B}T}}+e^{\frac{-\omega}{2k_{B}T}}\ $$
and the non zero density matrix elements are given by
\begin{footnotesize}\begin{align} 
\rho_{11}&=e^{\frac{3\omega}{2k_{B}T}};~ ~~\rho_{88}=e^{\frac{-3\omega}{2k_{B}T}} \nonumber \\
\rho_{22}&=\rho_{44}=\frac{1}{4}\left[ e^{\frac{(\frac{\omega}{2}-\sqrt{2}\Omega)}{k_{B}T}}+e^{\frac{(\frac{\omega}{2}+\sqrt{2}\Omega)}{k_{B}T}}+2 e^{\frac{\omega}{2k_{B}T}}\right];~~\rho_{33}=\frac{1}{2}\left[ e^{\frac{(\frac{\omega}{2}-\sqrt{2}\Omega)}{k_{B}T}}+e^{\frac{(\frac{\omega}{2}+\sqrt{2}\Omega)}{k_{B}T}}\right] \nonumber \\
\rho_{55}&=\rho_{77}=\frac{1}{4}\left[ e^{-{(\frac{\omega}{2}-\sqrt{2}\Omega)}{k_{B}T}}+e^{-\frac{(\frac{\omega}{2}+\sqrt{2}\Omega)}{k_{B}T}}+2 e^{-\frac{\omega}{2k_{B}T}}\right];~\rho_{66}=\frac{1}{2}\left[ e^{-\frac{(\frac{\omega}{2}-\sqrt{2}\Omega)}{k_{B}T}}+e^{-\frac{(\frac{\omega}{2}+\sqrt{2}\Omega)}{k_{B}T}}\right] \nonumber \\
\rho_{23}&=\rho_{34}=\frac{1}{2\sqrt{2}}\left[ e^{\frac{(\frac{\omega}{2}-\sqrt{2}\Omega)}{k_{B}T}}-e^{\frac{(\frac{\omega}{2}+\sqrt{2}\Omega)}{k_{B}T}}\right];~~\rho_{24}=\frac{1}{4}\left[ e^{\frac{(\frac{\omega}{2}-\sqrt{2}\Omega)}{k_{B}T}}+e^{\frac{(\frac{\omega}{2}+\sqrt{2}\Omega)}{k_{B}T}}-2 e^{\frac{\omega}{2k_{B}T}}\right]\nonumber \\
\rho_{56}&=\rho_{67}=\frac{1}{2\sqrt{2}}\left[ e^{-\frac{(\frac{\omega}{2}+\sqrt{2}\Omega)}{k_{B}T}}-e^{-\frac{(\frac{\omega}{2}-\sqrt{2}\Omega)}{k_{B}T}}\right];~\rho_{57}=\frac{1}{4}\left[ e^{-\frac{(\frac{\omega}{2}+\sqrt{2}\Omega)}{k_{B}T}}+e^{-\frac{(\frac{\omega}{2}-\sqrt{2}\Omega)}{k_{B}T}}-2 e^{-\frac{\omega}{2k_{B}T}}\right].\nonumber \\
\end{align}
\end{footnotesize}
It is well known that a mixed state $ \rho $ is separable if it can be expressed as a convex sum of three pure states, $|\psi \rangle$ = $|\phi_{A} \rangle \otimes |\phi_{B} \rangle \otimes |\phi_{C} \rangle $, otherwise it is called an entangled state. 
 In this case, at high temperature $ \beta \rightarrow 0$, the density matrix $ \rho_{ABC} $ of the system reduces to,
\begin{equation}
  \rho =\dfrac{1}{8}\begin{bmatrix}
1 & 0 & 0 & 0 & 0  & 0 & 0 & 0 & \\
0 & 1 & 0 & 0 & 0  & 0  & 0 & 0  \\
0 & 0 & 1 & 0 & 0 & 0 & 0 & 0  \\
0 & 0 & 0 & 1 & 0 & 0 & 0 &  0  \\
0 & 0 & 0& 0 & 1 & 0 & 0 & 0 \\
0 & 0 & 0 & 0 & 0 & 1 & 0 & 0 \\
0 & 0 & 0 & 0 & 0 & 0 & 1 & 0\\
0 & 0 & 0  & 0 & 0 & 0 & 0 & 1 \\
\end{bmatrix} 
\label{separable_density}
\end{equation} 
 which can further be expressed as a convex combination of three pure states,
\begin{footnotesize}
\begin{eqnarray}
  \rho =\dfrac{1}{8}\begin{pmatrix}
1 & 0 \\
0 & 0 \\
\end{pmatrix} \otimes \begin{pmatrix}
1 & 0 \\
0 & 0 \\
\end{pmatrix} \otimes \begin{pmatrix}
1 & 0 \\
0 & 0 \\
\end{pmatrix} +....+\dfrac{1}{8}\begin{pmatrix}
0 & 0 \\
0 & 1 \\
\end{pmatrix} \otimes \begin{pmatrix}
0 & 0 \\
0 & 1 \\
\end{pmatrix} \otimes \begin{pmatrix}
0 & 0 \\
0 & 1 \\
\end{pmatrix}
\end{eqnarray} 
\end{footnotesize}

From the above description, one observes that the system at high temperature is perfectly separable.  However, for intermediate temperatures, the system is in a mixed state and we investigate the entanglement properties of such system, from a study of  the pairwise concurrence  $ C(\rho_{ij}) $ and the quantum discord  $ D(\rho_{ij}) $.

\subsection[Results]{Numerical Results}
In this section, we study the pairwise concurrence $ C(\rho_{ij}) $ and discord $  D(\rho_{ij}) $ as functions of the ratio of dipole-dipole coupling strength $ \Omega $  to atomic transition frequency.   In addition, the influence
of atomic transition frequency $ \omega $ and the temperature $ k_{B}T $ on the discord and concurrence of the system are also discussed.
\begin{figure}[H]
\centerline{\includegraphics[width=16.5 cm,height=7.0 cm]{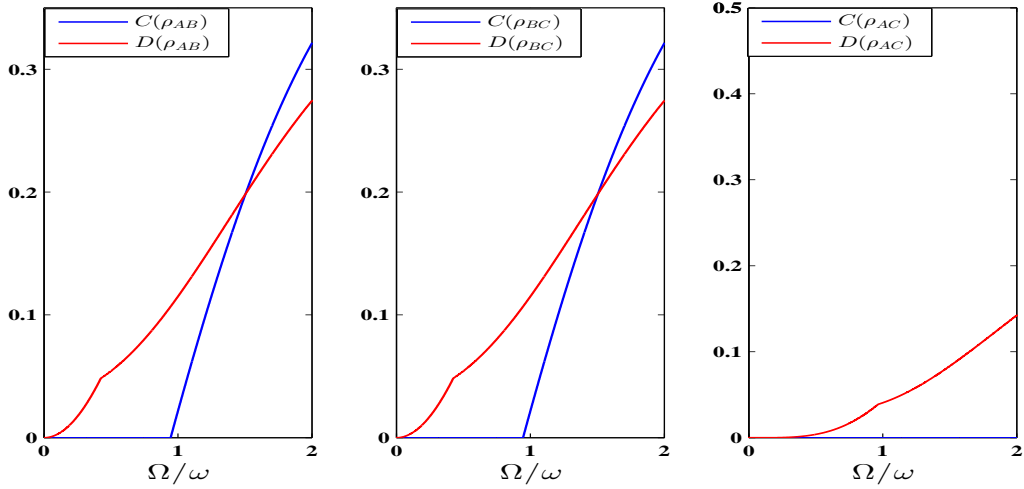}}
\caption[]{Line configuration: (Color online) Pairwise $ C(\rho_{ij}) $, $ D(\rho_{ij}) $ as a function of $ \Omega/\omega $  for $ k_{B}T/\omega =1$.}
\label{fig:A}
\end{figure}
Figure \ref{fig:A} shows the variation of the pairwise concurrence $ C(\rho_{ij}) $ and discord $ D(\rho_{ij}) $ as a function of $ \Omega/\omega $ for the temperature $ k_{B}T/\omega =1 $.  

 From the figure, it is clearly seen that $ C(\rho_{ij}) $ and $D(\rho_{ij})$ are zero when there is no coupling between the atoms.  With increasing $ \Omega/\omega $, it is observed that $ C(\rho_{AB})=C(\rho_{BC}) $ and $ D(\rho_{AB})=D(\rho_{BC}) $ where as $ C(\rho_{AC}) =0 $.  But $ D(\rho_{AC}) $ is non zero which implies that quantum correlations are still present even though end atoms A and C are not directly dipole coupled. 
 We also note that the pairwise concurrence $ C(\rho_{AB}),C(\rho_{BC}) $ has non zero values only when $ \Omega \geq k_{B}T $.  In addition, it is observed that the quantum discord $ D(\rho_{AB})(D(\rho_{BC})) $ dominates over  the concurrence $ C(\rho_{AB})(C(\rho_{BC})) $ when  the condition $ \Omega/\omega \leq 1.5 $ is met.  In the other regime, i.e., for $ \Omega/\omega > 1.5 $, concurrence dominates over the corresponding discord.  This clearly shows that one can not write a simple relation between the concurrence and the discord for any subsystem comprising of one pair of qubits.  This result is in agreement with that of\cite{Luo}, in which they have considered a system of two qubits. 
\begin{figure}[H]
\centerline{\includegraphics[width=16.5 cm,height=7.0 cm]{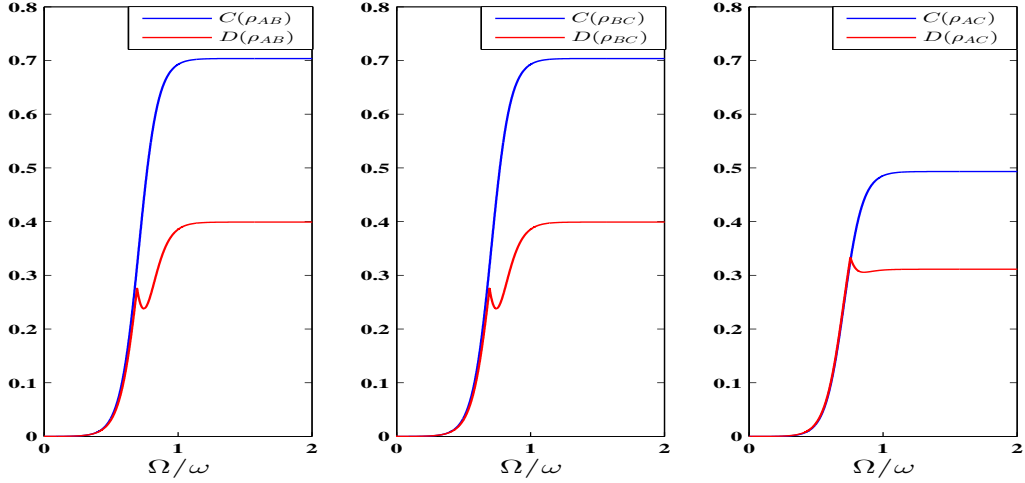}}
\caption[]{Line configuration: (Color online) Pairwise $ C(\rho_{ij}) $, $ D(\rho_{ij}) $ as a function of $ \Omega/\omega $  for $ k_{B}T/\omega  = 0.1$.}
\label{fig:B}
\end{figure}

\begin{figure}[H]
\centerline{\includegraphics[width=16.5 cm,height=7.0 cm]{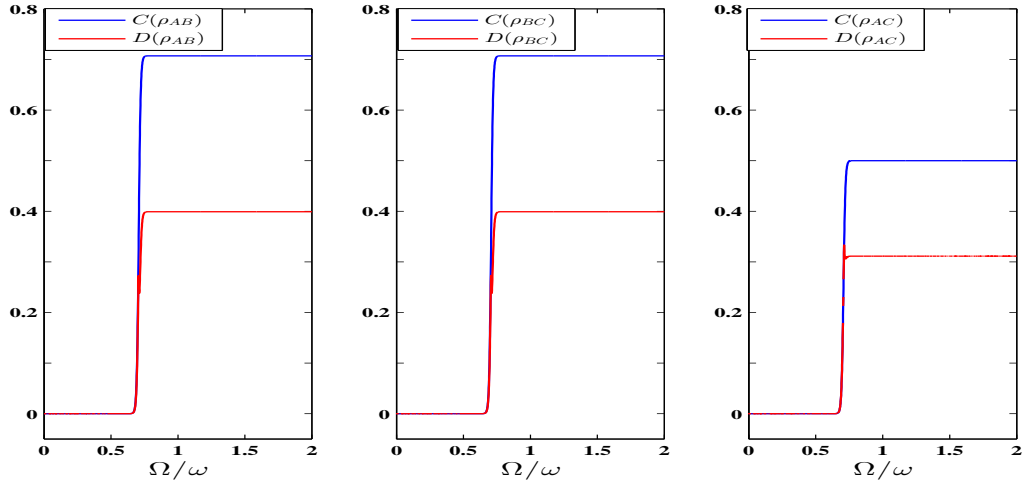}}
\caption[]{Line configuration: (Color online) Pairwise $ C(\rho_{ij}) $, $ D(\rho_{ij}) $ as a function of $ \Omega/\omega $  for $ k_{B}T/\omega  = 0.01$.}
\label{fig:BB}
\end{figure}
\begin{figure}[H]
\centerline{\includegraphics[width=16.5 cm,height=6.0 cm]{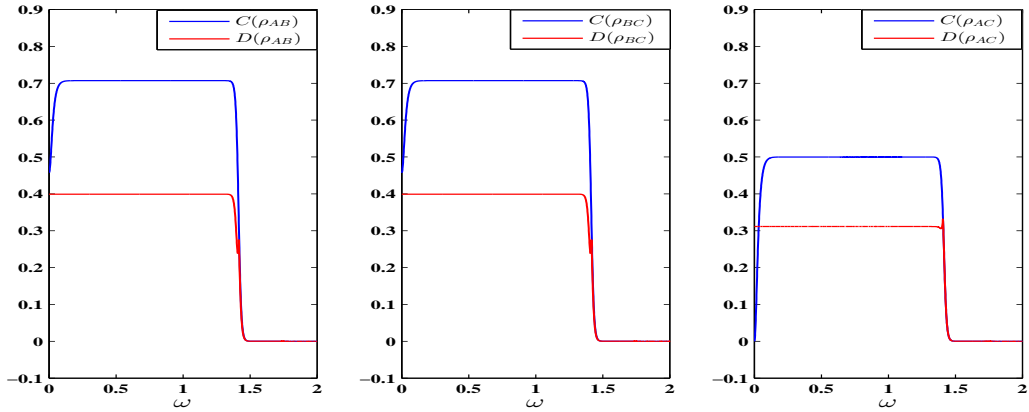}}
\caption[]{Line configuration: (Color online) Pairwise $ C(\rho_{ij}) $, $ D(\rho_{ij}) $ as a function of $ \omega $ for $ k_{B}T=0.01$ and $\Omega=1$.}
\label{fig:C}
\end{figure}

In the lower temperature regime, the concurrence  $C(\rho_{ij})$ and the quantum correlations $D(\rho_{ij})$ are greatly enhanced as shown in figures \ref{fig:B} and \ref{fig:BB}.  Here the figures \ref{fig:B} and \ref{fig:BB} show the variation of $C(\rho_{ij})$ and $D(\rho_{ij})$ as a function of $ \omega/\Omega $ for $k_{B}T/\omega =0.1$ and $k_{B}T/\omega =0.01$, respectively.  For $k_{B}T/\omega=0.1$, it is observed that the pairwise concurrence $ C(\rho_{AB})= C(\rho_{BC})$ and $ D(\rho_{AB})= D(\rho_{BC})$ whereas $ C(\rho_{AC})$ attains a non zero value.    
 We can thus deduce from a careful study of figures.\ref{fig:B} and \ref{fig:BB} that the quantum correlations $D(\rho_{ij})$ and pairwise entanglement $ C(\rho_{ij}) $ tend to increase with increasing dipole-dipole coupling strength, and attain a maximum/saturation value.  Further decrease in temperature shows a very interesting feature in both these quantities.  It is clearly seen that with a decrease of temperature, the curves of $ C(\rho_{ij}) $ and $D(\rho_{ij})$ plotted as a function of $ \Omega/\omega $ attain sharper edges, with all the kinks removed, thus  resembling the shape of a switch, which in this context is referred to as a  correlation /quantum switch\cite{Qi}.  This behaviour is much better elucidated when  the quantities $ C(\rho_{ij}) $ and $D(\rho_{ij})$ are plotted as functions of atomic transition frequency $ \omega $ in the lower temperature regime ($ k_{B}T=0.01$), which are shown in figure \ref{fig:C}.  From the results presented here for different temperatures and dipole coupling strengths, it is clear that these quantum correlation switches can be constructed by properly tuning the temperature and dipole coupling strengths.    

\begin{figure}[H]
\centerline{\includegraphics[width=14.5 cm,height=6.0 cm]{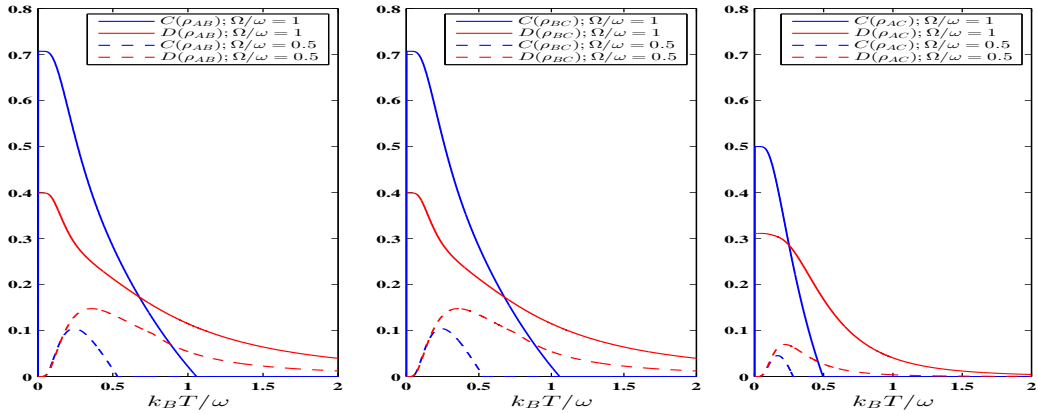}}
\caption[]{Line configuration: (Color online) Pairwise $ C(\rho_{ij}) $, $ D(\rho_{ij}) $ as a function of $ k_{B}T/ \omega$ for $ \Omega/\omega = 1$ and $ 0.5 $.}
\label{fig:D}
\end{figure}

In figure \ref{fig:D}, we study the pairwise $ C(\rho_{ij}) $ and $ D(\rho_{ij}) $ as a function of $ k_{B}T/ \omega$ for different values of dipole-dipole coupling strength ($\Omega/\omega = 1, 0.5 $).  It is clearly seen that the pairwise quantum discord $ D(\rho_{ij}) $ has non zero values at $ k_{B}T=0 $.  This is in contrast to the property of pairwise entanglement $ C(\rho_{ij}) $.  The concurrence $ C(\rho_{ij}) $ vanishes  at $ k_{B}T=0 $, and then attains non-zero values as the temperature is raised and reaches a maximum value after which it falls to zero very rapidly at a  specific temperature, namely when $ k_{B}T= \Omega $.  But the discord $ D(\rho_{ij}) $  continues to remain non-zero and vanishes at much higher temperatures.  This indicates the robustness of quantum discord $ D(\rho_{ij}) $ against the concurrence as the temperature is increased.  For higher values of dipole coupling strengths ($\Omega/\omega$) also, similar features are observed.  To conclude, quantum correlations may be enhanced by  lowering the temperature and/or by increasing the dipole dipole coupling strength, when the atoms are arranged on a line.

\section[Loop configuration]{The Quantum correlations in Loop configuration}
In this section, we present a discussion of the results that are obtained on  the thermal entanglement properties of three atoms arranged in a loop configuration.  In this case, we observe that the reduced density matrices $ \rho_{AB} $, $ \rho_{BC} $ and $ \rho_{AC} $ have different entanglement characteristics, though all of them have identical X-state form. 
 
\begin{figure}[H]
\centerline{\includegraphics[width=16.5 cm,height=6.0 cm]{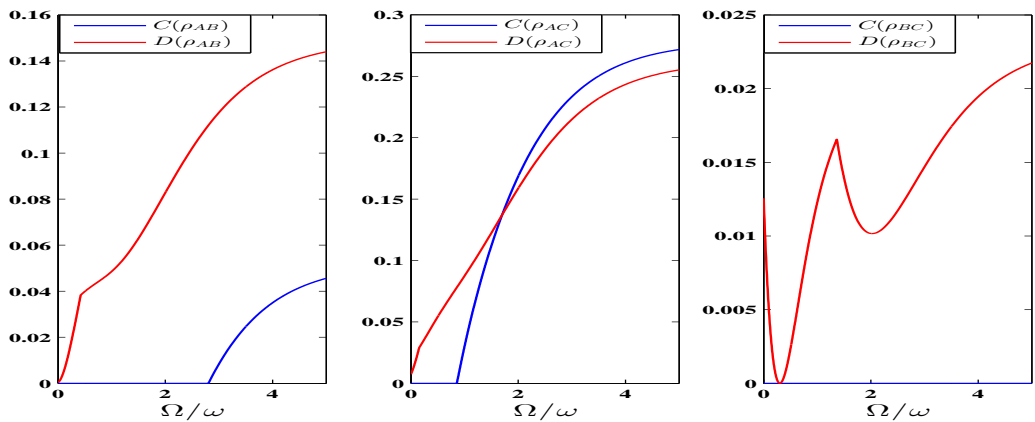}}
\caption[]{Loop configuration: (Color online) Pairwise $ C(\rho_{ij}) $, $ D(\rho_{ij}) $ as a function of $\Omega/\omega $ for $ k_{B}T/\omega = 1 $.}
\label{fig:E}
\end{figure}

A comparison between the quantum discord $ D(\rho_{ij}) $ and the concurrence $ C(\rho_{ij}) $ for $ k_{B}T/\omega =1.0 $ is displayed in figure\ref{fig:E}.  From the figure, it is clearly seen that reduced density matrix $ \rho_{BC} $ is separable and contains zero entanglement for all values of dipole-dipole coupling $ \Omega $, whereas the other reduced density matrices ($ C(\rho_{AB})\neq C(\rho_{AC}) $) have non-zero entanglement.  This surprising yet remarkable feature has been discussed in the literature \cite{Sab,star} for certain pure states.
Similarly, one notes that the quantum correlations $ D(\rho_{AB})\neq D(\rho_{AC})\neq D(\rho_{BC}) $.   

\begin{figure}[H]
\centerline{\includegraphics[width=16.5 cm,height=6.0 cm]{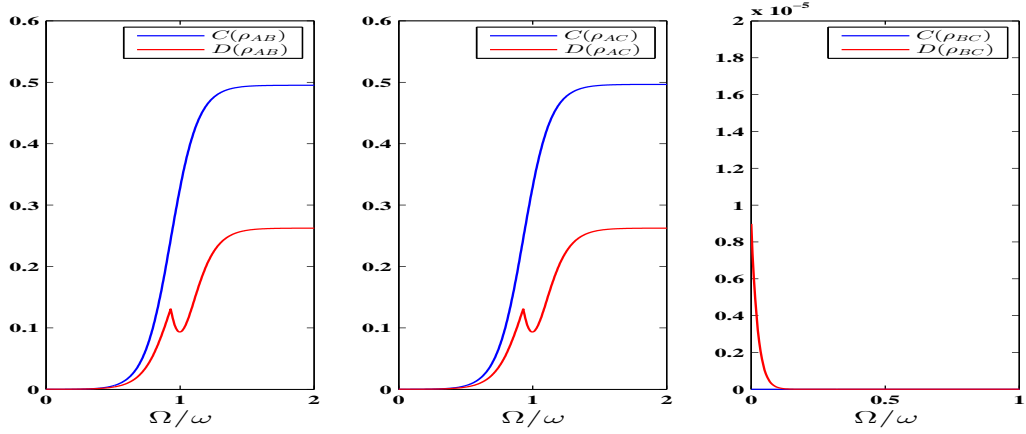}}
\caption[]{Loop configuration: (Color online) Pairwise $ C(\rho_{ij}) $, $ D(\rho_{ij}) $ as a function of $\Omega/\omega $ for $ k_{B}T/\omega = 0.1 $.}
\label{fig:F}
\end{figure}

\begin{figure}[H]
\centerline{\includegraphics[width=16 cm,height=6.0 cm]{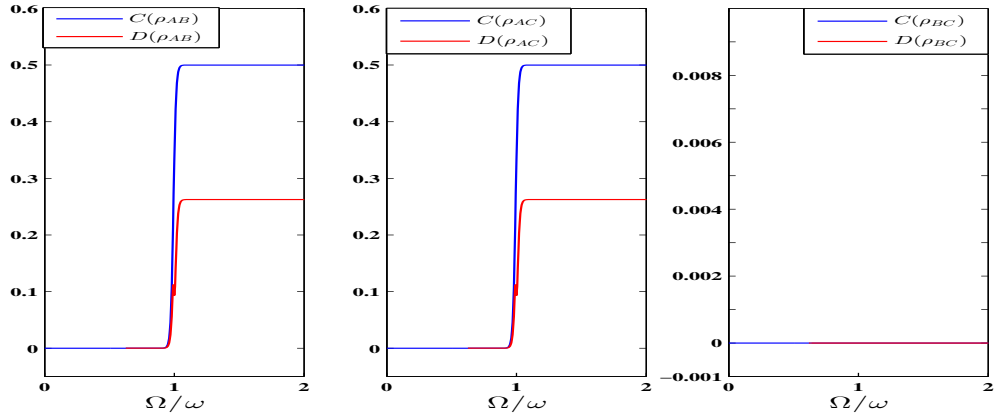}}
\caption[]{Loop configuration: (Color online) Pairwise $ C(\rho_{ij}) $, $ D(\rho_{ij}) $ as a function of $\Omega/\omega $ for $ k_{B}T/\omega = 0.01 $.}
\label{fig:FF}
\end{figure}

The same quantities as figure \ref{fig:E} for lower temperatures, viz.,  $ k_{B}T/\omega=0.1 $ and $ k_{B}T/\omega =0.01 $  are illustrated in figures \ref{fig:F} and \ref{fig:FF}  respectively.   As the temperature is lowered from $ k_{B}T/\omega =1 $ to 0.1 and 0.01, there is a significant increase in the  quantum discord $ D(\rho_{AB}) $, $ D(\rho_{AC}) $  and the concurrence $ C(\rho_{AB}) $, $ C(\rho_{AC}) $.  However, the discord  $ D(\rho_{BC}) $, which had a non-zero value at a higher temperature shows a monotonic decrease with lowered temperatures and further, at sufficiently  lower temperature, it approaches zero.   

\begin{figure}[H]
\centerline{\includegraphics[width=14 cm,height=6.0 cm]{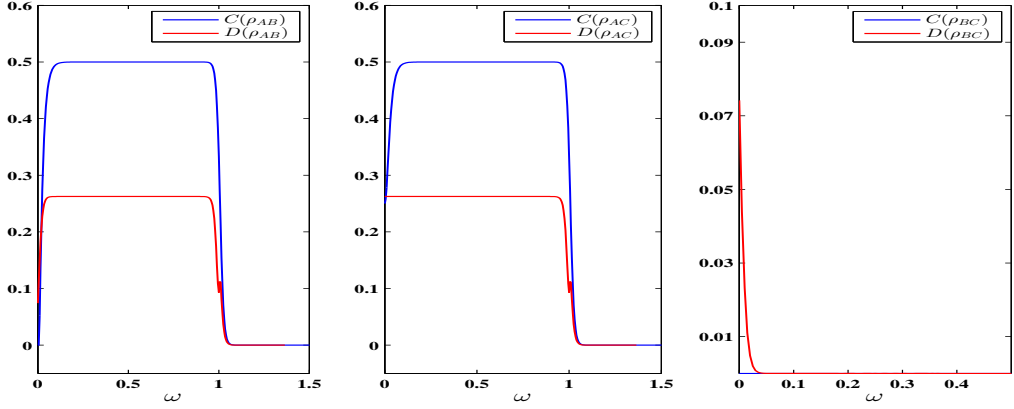}}
\caption[]{Loop configuration: (Color online) for Pairwise $ C(\rho_{ij}) $, $ D(\rho_{ij}) $ as a function of $ \omega $ $ k_{B}T=0.01 $ and $ \Omega = 1 $.}
\label{fig:G}
\end{figure}

The  interesting feature of  switch-like behaviour  at lower temperatures, exhibited by the atoms arranged in line configuration is also seen when they are arranged in a loop[cf fig. \ref{fig:G}].  However,  unlike in the line configuration, in the case of the loop configuration, only two of the correlations corresponding to $\rho_{AB}$ and  $\rho_{AC}$, show this switch-type feature,  while the third correlation (involving $\rho_{BC}$) is nearly zero, with only the discord having a small nonzero value at very small values of $\omega$, and quickly becoming zero with further increase in $\omega$,  with clear absence of the 'switch' feature.  It is believed that these quantum switches have important consequences in the implementation of quantum  gates in quantum computing. 

 \begin{figure}[H]
\centerline{\includegraphics[width=14.0 cm,height=6.0 cm]{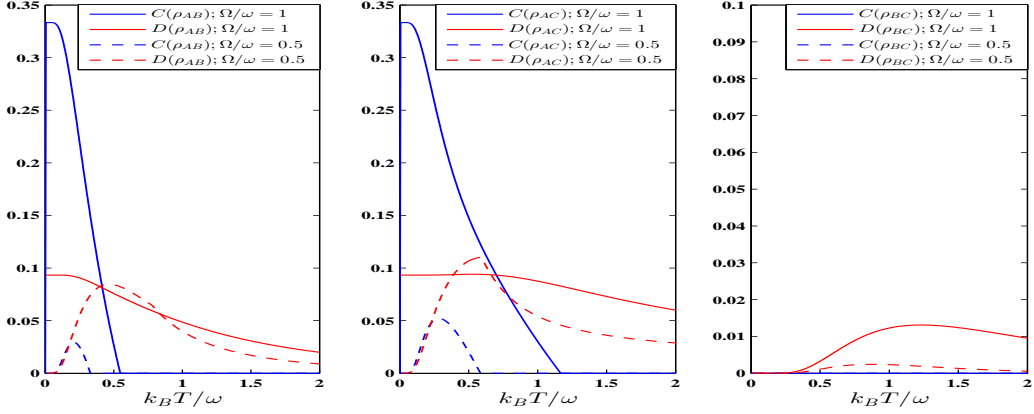}}
\caption[]{Loop configuration: (Color online) Pairwise $ C(\rho_{ij}) $, $ D(\rho_{ij})$ as a function of $ k_{B}T/\omega $ for different values of $ \Omega/\omega = 1, 0.5$.}
\label{fig:I}
\end{figure}
  In figure \ref{fig:I}, all the pairs of quantum correlations are plotted for typical values of dipole coupling strength $\Omega$ as a function of temperature.  As seen from figure \ref{fig:I}, the two concurrences $C(\rho_{AB})$ and $C(\rho_{AC})$ start from 0 at zero temperature, attain a maximum value at certain temperature and then decrease to zero with further increase in the temperature, while the corresponding discords $D(\rho_{AB})$ and $D(\rho_{AC})$ both peak  at zero temperature and decrease as the temperature is increased, for all values of the coupling strength $\Omega$, that are considered here.  Further, these concurrences rapidly diminish while the quantum discords still persist, with a much slower decay.  In general, both types of quantum correlations are seen to decay with increase in temperature due to thermal relaxation effects.  
 In summary, it is seen that the quantum correlations increase with increase in the dipole-coupling strength and decrease as the temperature is increased, which as is already discussed is due to thermal relaxation effects.  
     
\section{CONCLUSION}

The temperature dependent behaviour of the two quantum correlations, concurrence and discord, for all possible bipartite subsystems of the three coupled atoms in both line and loop configuration is studied in detail.  In both the line and loop configurations, it is seen that the quantum correlations increase with increase in the ratio of dipole-coupling strength to atomic transition frequency and decrease as the ratio of temperature to atomic transition frequency is increased, the latter of which is attributed to thermal relaxation.  In particular when the atoms are in loop configuration, certain interesting features are observed which are in agreement with the earlier reported studies on similar systems.  It is worth mentioning here that D$\ddot{u}$r \cite{Dur} has shown that the entanglement properties of a qubit with its neighbours are not simply determined by the mere number of its entangled neighbours but also by the properties of these entangled neighbours.  In other words, the presence of entanglement or classical correlations on certain pairs of qubits may imply correlations on other pairs, if these are in any way connected to the qubits.  These features very well substantiate our results on the entanglement properties, inferred from a study of the quantum correlations, of atoms arranged in a loop configuration.
 

\begin{thebibliography}{99}
 \bibitem{Ein}{A. Einstein, B. Podolsky, and N. Rosen, Phys. Rev. 4\textbf{7}, 777 (1935).}
\bibitem{Sch}{E. Schr\"{o}dinger, Naturwissenschaften 23, 807; ibid 823; ibid 844 (1935).}
\bibitem{Dik}{D. Bouwmeester, J. W. Pan, K. Mattle, M. Eibl, H. Weinfurter and A. Zeilinger, Nature (London) \textbf{390}, 575 (1997).}
\bibitem{Ekert}{A. Ekert, Phys. Rev. Lett. \textbf{67}, 661 (1991).}
\bibitem{CH}{C. H. Bennett and S. J. Wiesner, Phys. Rev. Lett. \textbf{69}, 2881 (1992).}
\bibitem{Nielson}{M. A. Nielsen and  I. L. Chuang, \textit{ Quantum Computation and Quantum Information}, Cambridge University Press, Cambridge (2000).}
\bibitem{Olliver}{H. Ollivier and W. H. Zurek, Phys. Rev. Lett. \textbf{88}, 017901(2001).}
\bibitem{Zurek}{ H. Zurek, Ann. Phys. (Berlin) \textbf{9}, 855(2000).}
\bibitem{Henderson}{L. Henderson and V. Vedral, J. Phys. A \textbf{34}, 6899(2001).}
\bibitem{Vedral}{V. Vedral, Phys. Rev. Lett. \textbf{90}, 050401(2003).}
\bibitem{wer}{T. Werlang, C. Trippe,  G. A. P. Ribeiro, G. Rigolin, Phys. Rev. Lett. \textbf{105}, 095702 (2010).}
 \bibitem{Auc}{R. Auccaise, L. C. C$\acute{e}$leri, D. O. Soares-Pinto, E. R. deAzevedo, J. Maziero, A. M. Souza, T. J. Bonagamba, R. S. Sarthour, I. S. Oliveira, and R. M. Serra, Phys. Rev. Lett. \textbf{107}, 140403 (2011).}
 \bibitem{chiran}{H. Singh, T. Chakraborty, P. K. Panigrahi and C. Mitra, Quantum Inf Process \textbf{14}, 951 (2015).}
 \bibitem{Dill}{R. Dillenschneider, Phys. Rev. B \textbf{78}, 224413 (2008).}
 \bibitem{Werlang}{T. Werlang, G. Rigolin, Phys. Rev. A \textbf{81}, 044101 (2010).}
 \bibitem{Bennett}{ C. H. Bennett, G. Brassard, C. Crepeau, R. Jozsa, A. Peres and W. K. Wootters, Phys. Rev. Lett. \textbf{70}, 1895 (1993).}
\bibitem{Bose}{S. Bose, V. Vedral and P. L. Knight, Phys. Rev. A \textbf{57}, 822 (1998).}
\bibitem{Ban}{M. Ban, J. Opt. B: Quantum Semiclass. Opt. \textbf{1}, L9 (1999).}
\bibitem{Pati}{A. K. Pati, P. Parashar and P. Agrawal, Phys. Rev. A \textbf{72}, 012329 (2005).}
  \bibitem{r4}{S. Hill and W. K. Wootters, Phys. Rev. Lett. \textbf{78}, 5022 (1997).} 
\bibitem{r5}{W. K. Wootters, Phys. Rev. Lett. \textbf{80}, 2245 (1998).}
\bibitem{rev}{V. Vedral, Rev. Mod. Phys. \textbf{74}, 197(2002).}
\bibitem{Berry}{B. Groisman, S. Popescu and A. Winter, Phys. Rev. A \textbf{72}, 032317(2005).}
\bibitem{Holevo}{A. S. Holevo, \textit{Probability and Statistical Aspects of Quantum Theory}, MTsNMO, Moscow (2003).}
\bibitem{Luo}{S. Luo, Phys. Rev. A \textbf{77}, 042303 (2008).}
\bibitem{Qi}{Q. Wei, S. Kais and Y. P. Chen, J. Chem. Phys. \textbf{132}, 121104 (2010 ).} 
\bibitem{Sab}{C. Sab$\acute{i}$n  and G. Garc$\acute{i}$a-Alcaine, Eur. Phys. J. D \textbf{48}, 435(2008).}
\bibitem{star}{M. Plesch and  V. Bu$\check{z}$ek, Phys. Rev. A \textbf{67}, 012322 (2003).}
\bibitem{Dur}{W. D$ \ddot{u} $r, Phys. Rev. A \textbf{63}, 020303 (2001).}


\end{thebibliography}
\end{document}